\documentclass[11pt]{article}
\usepackage{amsfonts,amsmath,amssymb,bm,hyperref,graphicx}
\usepackage{amsfonts,amsmath,amssymb,bm,graphicx}         
\makeatletter
\newcommand\figcaption{\def\@captype{figure}\caption}
\newcommand\tabcaption{\def\@captype{table}\caption}
\makeatother \textheight 25 true cm \textwidth 16 true cm \topmargin
-2.5cm \oddsidemargin 0 true cm
\title{Restudy on Dark Matter Time-Evolution in the Littlest Higgs model with T-parity}
\author{{Qing-Peng Qiao, ~Jian Tang, ~Xue-Qian Li}\\
{\small \it Department of Physics, Nankai University, TianJin
300071,China}}
\date{}
\begin{document}
\maketitle
{\bf Abstract}\\

Following Refs. \cite{Hubisz:2004,Andreas:2006}, in the Littlest
Higgs model (LHM), the heavy photon is supposed to be a possible
dark matter candidate and its relic abundance of the heavy photon is
estimated in terms of the Boltzman-Lee-Weinberg time-evolution
equation. The effects of the T-parity violation is also considered.
Our calculations show that when Higgs mass $M_H$ taken to be 300 GeV
and don't consider T-parity violation, only two narrow ranges
$133<M_{A_{H}}<135$ GeV and $167<M_{A_{H}}<169$ GeV are tolerable
with the current astrophysical observation and if
$135<M_{A_{H}}<167$ GeV, there must at least exist another species
of heavy particle contributing to the cold dark matter. As long as
the T-parity can be violated, the heavy photon can decay into
regular standard model particles and would affect the dark matter
abundance in the universe, we discuss the constraint on the T-parity
violation parameter based on the present data. Direct detection
prospects are also discussed in some detail.\\


\section*{I. Introduction} \hspace{0mm}\vspace{2mm}\\

The Littlest Higgs model~\cite{Arkani:2002} has been proposed for
solving the hierarchy problem of Standard Model (SM) recently.
Unfortunately, the original Littlest Higgs model still suffers from
severe constraints from precision electroweak fits and the
fine-tuning of the Higgs boson mass is necessary. To avoid the
difficulty, a discrete $Z_{2}$ symmetry named as
"T-parity"~\cite{Cheng:2003} (just as R-parity in SUSY) is
introduced to the model. This symmetry in the little Higgs models
endow SM-like particles T-even parity while all heavy gauge bosons
and scalar triplets are T-odd. Therefore, it is completely natural
to see that the problem in the model can be avoided. An interesting
feature of the Littlest Higgs Model with T-parity is that the T-odd
particles need to be pair-produced and will cascade down to the
lightest T-odd particle (LTP) which means that the LTP is guaranteed
to be stable. At the same time, one of the most fundamental problems
in cosmology and particle physics today is {\it what is the nature
of the Dark Matter}. A flood of discussions point out that the Dark
Matter should have the following characters -- non-luminous,
non-baryonic, non-relativistic, and electrically neutral \cite
{Rees:1986ma,Roszkowski:1991ng,Roszkowski:1999ts,Primack:2001ia,Chen:2003bn,Ma:2004nw}.\\

The heavy photon in the Littlest Higgs model indeed meets all
criteria, thus it can be an ideal candidate of cold dark matter
(CDM). The relic abundance of the dark matter in the Littlest Higgs
model with T-parity in the thermal relic scenario has already been
evaluated by Hubisz and Birkedal~\cite{Hubisz:2004,Andreas:2006}. To
be consistent with the  observation on CDM, the mass of of heavy
photon is estimated around a few hundred GeV.\\

In this work, following Refs.\cite{Hubisz:2004,Andreas:2006}, we
re-evaluate the abundance of heavy photons in our universe by means
of the Boltzman-Lee-Weinberg equation where the T-parity violation
effects are taken into account. As a reasonable consideration we set
the mass of $A_{H}$ in the range: 100 GeV $\leq M_{A_{H}} \leq$ 300
GeV . We find that a rather narrow window for the mass of heavy
photon meet the requirements of the observation, outside of which
other kinds of dark matter particles are necessary. The paper is
organized as follows: after this introduction, we formulate the time
evolution of the dark matter in the Littlest Higgs model with and
without T-parity violation in section II and then we present our
numerical results. Direct detection prospects are discussed in
section III. The section IV is devoted to our brief
conclusion and discussion.\\

\section*{II. The time evolution of the dark matter}\hspace{0mm}\vspace{2mm}\\

The time evolution of the dark matter is described by the
Boltzman-Lee-Weinberg time-evolution equation \cite{Lee:1977}, which
is the basis of study on the abundance of dark matter for a possible
candidate particle (heavy photon in this work) at the present stage:
\begin{eqnarray}
\frac{dn}{dt} = \frac{- 3\dot {R}}{R}n - < \sigma v
> n^2 + < \sigma v > n_0^2 ~,
\label{T-parity}
\end{eqnarray}
where $< \sigma v >$ is the average value of the heavy photon
pair-annihilation cross section times the relative velocity;R is the
cosmic scale factor and $\dot{R}/{R}$ is the Hubble constant H; n is
the number density of cold dark matter in the thermal bath and
$n_{0}$ is the equilibrium density:
\begin{eqnarray}
n_0 (T) = \frac{2}{(2\pi )^3}\int_0^\infty {4\pi p^2} \mbox{d}p
~[\exp \frac{(m^2 + p^2)^{\frac{1}{2}}}{kT} - 1]^{ - 1} ~,
\end{eqnarray}
the expression of $n_{0}$ is different to our former paper
\cite{Feng:2002}, where neutralino is supposed to be the cold dark
matter particle which is a fermion. Instead, $A_{H}$ we employ
here is a boson and obeys the Boson-Einstein statistics.\\

Obviously, in eq.(\ref{T-parity}), the T-parity is conserved, so
that the heavy photon as the lightest T-odd particle does not decay.
The evolution equation should be modified if we consider the
T-parity violation effects:
\begin{eqnarray}
\frac{dn}{dt} = \frac{- 3\dot {R}}{R}n - < \sigma v
> n^2 + < \sigma v > n_0^2 - \Gamma n ~,
\end{eqnarray}
where $\Gamma=1/\tau$ is the decay width of the $A_{H}$ which
decays into SM particles via T-parity violation. The $< \sigma v >$ means
the thermally averaged annihilation cross section.\\

In this work we only consider that the dark matter particles
annihilate into the SM particles in s-wave, thus under the
non-relativistic limit, the cross section $\sigma v$ is simply given as
\begin{eqnarray}
 \sigma v =\sigma v|_{_{WW}}+\sigma v|_{_{ZZ}}+\sigma
v|_{HH}+\sigma v|_{t\bar{t}}~.
\end{eqnarray}\\

Due to relatively large uncertainties of the present data, we
restrict ourselves to only evaluate the cross sections of the
2$\longleftrightarrow$ 2 processes which would be overwhelming over
the 2 to 3 modes, and the corresponding Feynman
diagrams are shown in Fig. \ref{feyn1}.
\\[\intextsep]
\begin{minipage}{\textwidth}
\centering
\includegraphics[height=5.0cm,angle=0]{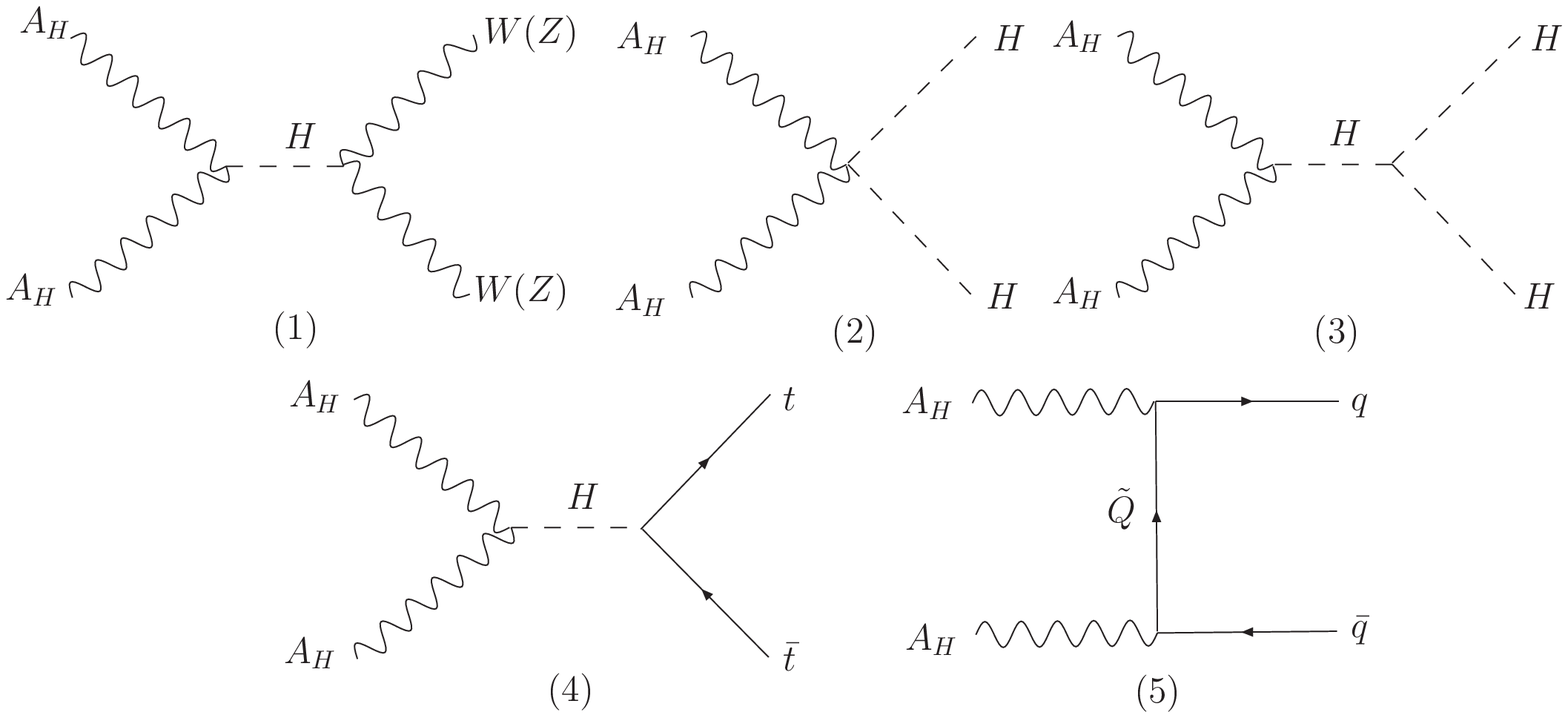} \figcaption{Feynman
diagrams for the annihilation of the dark matter ($A_{H}$) in the
 $2\leftrightarrow2$ processes. Diagrams (1) to (4) are those
which give the largest contributions to the annihilation $<\sigma
v>$ for the ranges of $M_{A{H}}$ and $M_{H}$ that we examine.
Diagram (5) contribution to the annihilation coefficient $< \sigma v
>$ is much smaller than the other four, so we ignore it in numerical
computation~\cite{Andreas:2006}.} \label{feyn1}

\end{minipage}

\newpage

For the readers' convenience, we present the vertices which we are
using in Table \ref{table1}, where the coupling $g^{'}$ =
$2\sqrt{\pi\alpha}/cos\theta_w$.
We can directly obtain $\sigma v$ of the diagrams shown in Fig. \ref{feyn1}.
The subscript "CM" here denotes the center-of-mass frame.\\
\begin{table}
\begin{center}
\begin{tabular}{|c|c|}\hline
\ $A_H^\mu A_H^\nu H$ &  $- \frac{i}{2}g^{'2}vg^{\mu \nu }$ \\
\hline \ $A_H^\mu A_H^\nu HH$ & $- \frac{i}{2}g^{'2}g^{\mu \nu }$
\\ \hline
\end{tabular}
\caption{The interaction vertices Littlest Higgs model that appear
in
 this paper.}
\end{center}
\label{table1}
\end{table}

\begin{eqnarray}
(\sigma v|_{_{WW}})_{CM} \thickapprox \displaystyle {{2\pi \alpha
^2} \over {3\cos ^4\theta _w }}\frac{M_{A_H}^2}{(4M_{A_H}^2 - M_H^2
)^2 + M_H^2 \Gamma _H^2 }(1 - \mu _w
+ \frac{3}{4}\mu _w^2 )\sqrt {1 - \mu _w }~,\\
 (\sigma v|_{_{ZZ}})_{CM}
\thickapprox \displaystyle{{\pi \alpha ^2} \over {3\cos ^4\theta _w
}}\frac{M_{A_H}^2}{(4M_{A_H}^2 - M_H^2 )^2 + M_H^2 \Gamma _H^2 }(1 -
\mu _z + \frac{3}{4}\mu _z^2 ) \sqrt {1 - \mu _z }~,
\end{eqnarray}
where $ \mu _i=M_i^2 /M_{A_{H}}^2 $ and $M_{A_{H}}$ is the mass of
$A_{H}$, $\theta_{w}$ is the SM Weinberg angle, $\alpha$ keeps the
meaning that it is in SM and $M_{H}$ is the mass of SM Higgs boson
and $\Gamma_{H}$ is the width of the Higgs. If $M_{A_{H}}> m_{t}$,
the annihilation of heavy photon pair into a pair of top quarks is
possible, and
\begin{eqnarray}
(\sigma v|_{t\bar {t}})_{CM} \thickapprox \displaystyle{{\pi \alpha
^2} \over {\cos ^4\theta _w }}\frac{M_{A_H}^2}{(4M_{A_H}^2 - M_H^2
)^2 + M_H^2 \Gamma _H^2 }\mu _t (1 - \mu _t )^{\frac{3}{2}}~.
\end{eqnarray}\\

If $M_{A_{H}} > M_{H}$, it also can annihilate into pairs of Higgs
bosons, with the following cross section~\cite{Andreas:2006},
\begin{eqnarray}
(\sigma v|_{HH})_{CM} \thickapprox \displaystyle{{\pi \alpha ^2
M_{A_H}^2} \over {2\cos ^4\theta _w }}(\frac{\mu _h (\mu _h +
8)}{8((4M_{A_H}^2 - M_H^2 )^2 + M_H^2 \Gamma _H^2 )} +
\frac{1}{24M_{A_H}^4})\sqrt {1 - \mu _H }~,
\end{eqnarray}\\

where the subscript "CM" denotes the center-of-mass. The dependence
of $\sigma v$ on the masses of Higgs and heavy photon is depicted in
Fig. \ref{contour},
\\[\intextsep]
\begin{minipage}{\textwidth}
\centering
\includegraphics[height=8cm,angle=0]{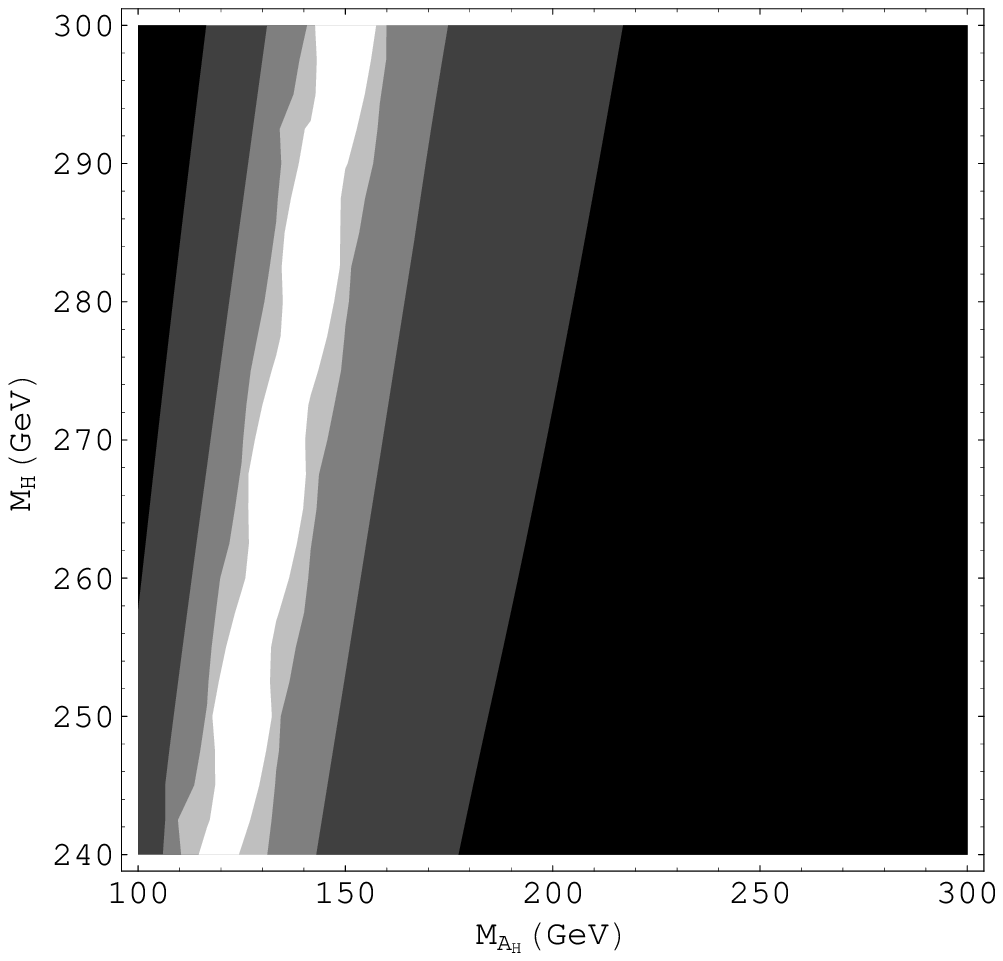}
\figcaption{This plot depicts the dependence of the  $<\sigma v>$
on the mass of Higgs and heavy photon. From the
lightest to the darkest regions, the magnitude of $<\sigma v>$ decreases
as shown in Fig.} \label{contour}
\end{minipage}\\
\\[\intextsep]

In the numerical computations, we take a few typical values for mass
of Higgs as $M_{H}$=240 GeV and $M_{H}$=300 GeV, as shown in Fig.
\ref{sigma}
,\\
\\[\intextsep]
\begin{minipage}[c]{0.5\textwidth}
\centering
\includegraphics[width=3in]{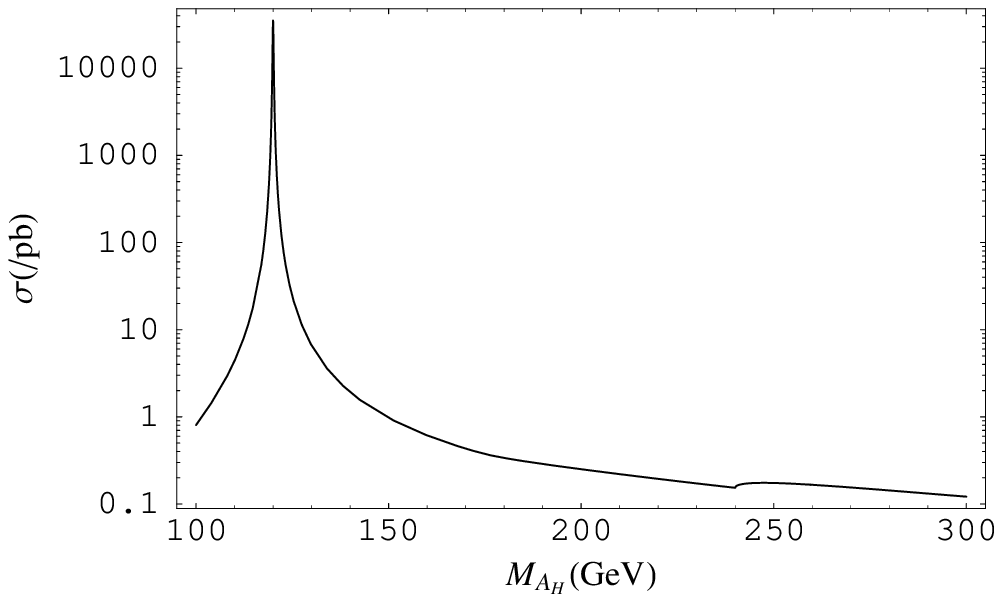}
\end{minipage}%
\begin{minipage}[c]{0.5\textwidth}
\centering
\includegraphics[width=3in]{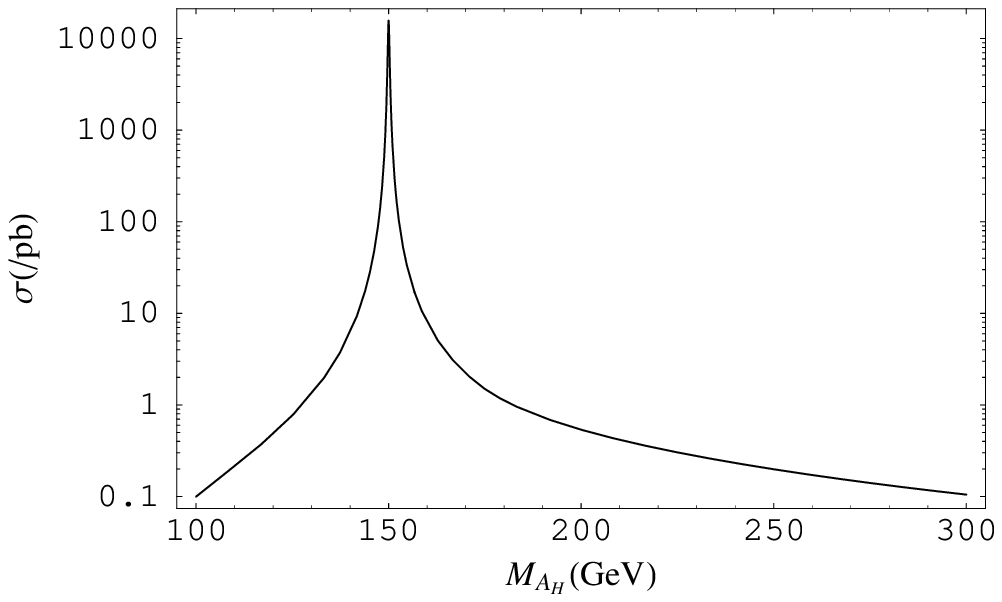}
\end{minipage}\\
\begin{minipage}{\textwidth}
\figcaption{the magnitude of $<\sigma v>$ with $M_{H}$=240 GeV
(left) and $M_{H}$=300 GeV (right).}\label{sigma}
\end{minipage}\\
\\[\intextsep]

It is noted that we have not considered the contributions of the
T-parity violation effects to $<\sigma v>$ yet. Later in this
work, we will show that to be consistent with the observational data
on dark matter density in our universe, if the heavy photon is the
only constituent of the dark matter, the T-parity violation cannot
be large. Thus when we evaluate $<\sigma v>$, the diagrams related
to the T-parity violation can be safely neglected. There is a sharp
peak in the figure which is due to the resonance effect
with $4M_{A_H}^2 - M_H^2\approx 0$.\\

For readers' convenience, let us briefly repeat the procedure for
solving Eq. (3). Following the literature, one needs to rewrite the
first term by means of the equation of state, according to the
Friedmann-Lema\^{i}tre equations
\begin{eqnarray}
H(t)= \frac{\dot {R}}{R} = (\frac{8\pi \rho G}{3} - \frac{\kappa
}{R^2} + \frac{\Lambda }{3})^{\frac{1}{2}}~,
\end{eqnarray}
where $H(t)$ is the Hubble parameter, $\rho$ is the energy density
and $\Lambda$ is the cosmological constant. Energy conservation
 $T_{;\nu }^{\mu \nu } = 0$ leads to another useful equation,
\begin{eqnarray}\dot {\rho } = - 3H(\rho + p)~.\end{eqnarray}
This equation can be written as $\dot {\rho } = - 3(1 + w)\rho
\dot {R} / R$ and is easily integrated to yield
\begin{eqnarray}
\rho \propto R^{ - 3(1 + w)}~,
\end{eqnarray}
where the state parameter $w=p/\rho$ is  1/3 at
the early universe dominated by radiation and 0 when matter dominates.\\

For the radiation-dominant universe, one has
\begin{eqnarray}
\rho =\frac{\pi ^2}{30}g^\ast (kT)^4~,
\end{eqnarray}
where $g^\ast = g_{_{B}} + \displaystyle\frac{7}{8}g_{_{F}}$ is the
 effective number of degrees of freedom. The $g_{_{B}}$ and
$g_{_{F}}$ denote the degrees of freedom for boson and fermion
respectively.\\

After the universe converted from radiation-dominance to
matter-dominance, the relation $RT=constant$ still
holds \cite{Lee:1977}, and
\begin{eqnarray}\rho = \beta T^3~,\end{eqnarray}
where constant $\beta$ can be fixed by the present energy density
\[
\rho _r = \beta T_r^3 = \Omega _m \rho _c~,
\]
where $T_r = 2.725$ K~\cite{PDG:2006} is the present microwave
background radiation temperature, $\Omega _m h^2 = 0.127$ is the
present-day density parameter for pressureless matter and the
critical density $\displaystyle\rho _c = \frac{3H^2}{8\pi G_N } =
1.05\times 10^{ - 5}h^2$ GeV$\cdot \textrm{cm}^{ -
3}$~\cite{PDG:2006}, thus
\begin{eqnarray}
\rho = \Omega
_m\rho _c(\frac{T}{T_r })^3~.
\end{eqnarray}\\

We have the modified evolution equation for the radiation-dominant
stage as
\begin{eqnarray} \frac{df}{dx} = (\frac{45}{4\pi ^3GN_F
})^{\frac{1}{2}}[\frac{M_{A_H } }{k} < \sigma v > (f^2 - f_0^2 ) +
\frac{\Gamma }{M_{A_H }^2 }\frac{f}{x^3}], \end{eqnarray} and the
other equation for the matter-dominant universe
\begin{eqnarray}\frac{df}{dx}
= (\frac{3T_r^3 }{8\pi G\rho _r })^{\frac{1}{2}}[(\frac{M_{A_H}
}{k})^{\frac{3}{2}} < \sigma v
> (f^2 - f_0^2 ) + \Gamma (\frac{k}{M_{A_H} })^{\frac{3}{2}}x^{
- \frac{5}{2}}f], \end{eqnarray} where \[ f(x) \equiv
\frac{n}{T^3}, \quad x \equiv \frac{kT}{M_{A_H} }, \quad f_0 =
\frac{n{ }_0}{T^3}={\displaystyle\frac{2}{(2\pi )^3}\int_0^\infty
{4\pi p^2} \mbox{d}p ~[\displaystyle\exp \frac{(m^2 +
p^2)^{\frac{1}{2}}}{kT} - 1]^{ - 1}}\frac{1}{T^{3}}~.
\]\\

In Fig. \ref{figure4}, we plot the dependence of $\rho/T^3$ on
temperature (the time). In order to illustrate our numerical
results, without losing generality we set the mass of Higgs as
$M_{H}=300$ GeV and will make more discussions in the last section.\\
\\[\intextsep]
\begin{minipage}{\textwidth}
\centering \centering
\includegraphics[height=9cm]{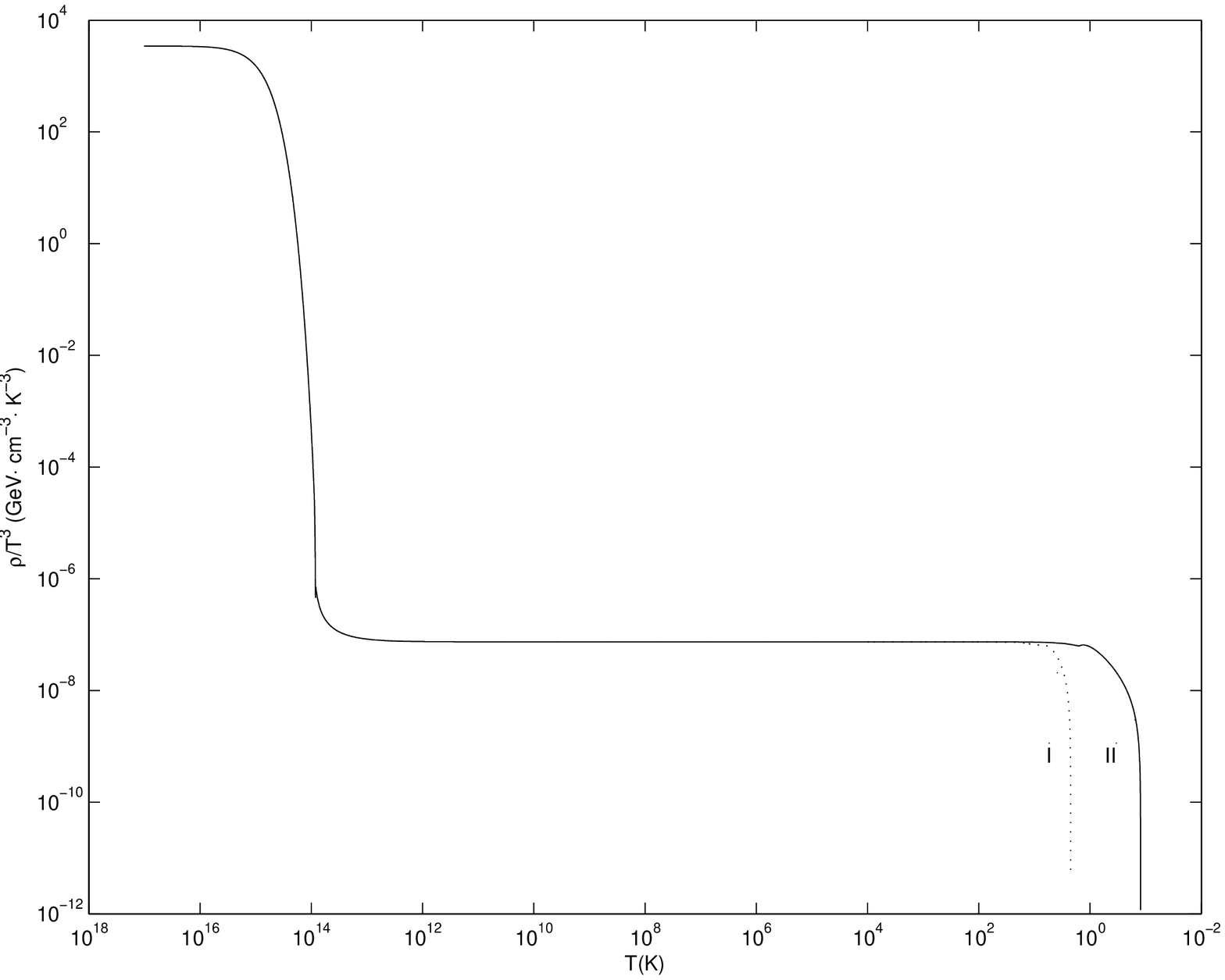}
\figcaption{ $\rho/T^{3}$ vs T. where two states of radiation and
matter
 are compared. Here $M_{H}=300$ GeV and $M_{A_{H}}=170$ GeV. In case I, as an
 approximation, we use radiation-dominance equation (15). The matter-dominance
 equation (16) is used in case II when temperature blow $10^{4}$ K.}\label{figure4}
\end{minipage}\\
\\[\intextsep]

The early universe was hot, dense and dominated by radiation, but as
the temperature gradually decreased, non-relativistic matter
eventually dominates the universe. Thus Eq.(15) and Eq.(16) describe
different stages of the temperature evolution and the solutions of
the two equations should be smoothly connected at a medium
temperature of about $10^{4}$ K. With this restriction we can obtain
our numerical results which are shown in Figs. \ref{figure4} $\sim$
Figs. \ref{figure7}.\\

By Fig. \ref{figure4}, one can notice that when temperature exceeds
$10^{16}$ K, $f(x)$ (or $\rho/T^{3}$) approaches to original
$f_{0}(x)$ (or $\rho_{0}$),namely the curve almost does not change
until the temperature drops to the decoupling temperature $T_{f}$.
As the temperature decreases continually the curve is flat again for
a long range until the temperature $T_{v}$
where T-parity violation applies, then the curve drops abruptly.\\

In this figure, for understanding the physics, we plot two curves.
The line I is a solution of Eq. (15) where one assumes the universe
is dominated by the radiation all the time, whereas the line II
corresponds to the case where the two solutions of Eq.(15) and
Eq.(16) are smoothly connected, namely the two stages of time
evolution are properly dealt with. One can notice that without
taking into account of the T-parity violation, line I and line II
are almost overlap, but when one considers the T-parity violation,
the line I drops earlier than the line II. It may influence the time
evolution of dark matter in the future (billion year later,
we suppose).\\

The Feynman diagrams which are responsible for the heavy photon
decaying into SM particles are shown in Fig. \ref{feyn2}, where the
vertex is T-parity violated.
\\[\intextsep]
\begin{minipage}{\textwidth}
\centering
\includegraphics[height=3cm,angle=0]{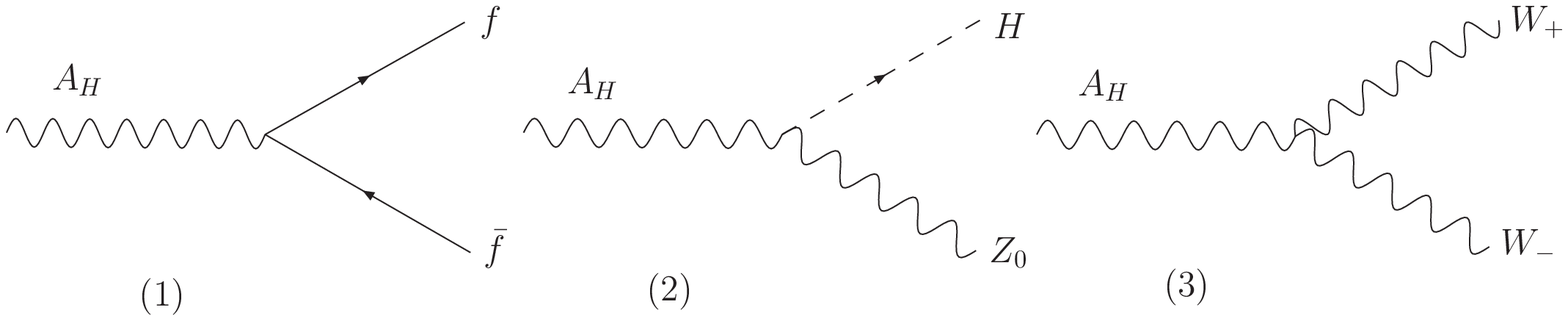} \figcaption{The Feynman
diagrams for the decay of the heavy photon $A_{H}$. Diagrams (2) to
(3) can be ignored since $Z_{0}$, $W_{\pm}$ and $H$ will decay
furthermore and the phase space is too narrow.  $f$ in diagram (1)
stands for all quarks, leptons except the top quark as long as
$A_{H}$ is set as 170 GeV in the numerical computations.}
\label{feyn2}
\end{minipage}\\
\\[\intextsep]

The decay amplitude of diagram (1) of Fig. \ref{feyn2} is
\begin{eqnarray}
M = \lambda_{f}\bar {u}(P_1 )\gamma ^\mu v(P_2 )\varepsilon _\mu
(K)~,
\end{eqnarray}
where $K$, $P_{1}$, $P_{2}$ denote the momenta of $A_{H}$, $f$ and
$\bar{f}$ respectively and $\lambda_{f}$ is the T-parity violation
parameter. The decay width of $A_{H}$ that decays to a pair of
$f\bar{f}$ is
\begin{eqnarray}
\Gamma _f =\frac{c_f \cdot \lambda _f^2 }{12\pi
}\displaystyle{{(M_{A_H }^2 + 2m_f^2 )(M_{A_H }^2 - 4m_f^2 )^{1 /
2}} \over {M_{A_H }^2 }}~.
\end{eqnarray}
Because $m_{f}\ll M_{A_H }$, we can ignore it safely in the
expression and obtain
\begin{eqnarray}
\Gamma _f = \frac{c_f \cdot \lambda _f^2 }{12\pi }M_{A_H }~,
\end{eqnarray}
where $c_{f}=3$ for quarks and $c_{f}=1$ for leptons. Finally, we
get the total decay width of $A_{H}$
\begin{eqnarray}
\Gamma _{total} = \sum\limits_f {\Gamma _f =
0.0796}M_{A_{H}}(5\lambda _1^2 + 2\lambda _2^2 ) =
0.0796M_{A_{H}}\lambda ^2~,
\end{eqnarray}
where $\lambda_{1}$, $\lambda_{2}$ denote the T-parity violation
parameters of quarks and leptons respectively. In general,
$\lambda_1$ and $\lambda_2$ may be different, but  the equation shows
that only their combination  applies in practice as an effective
parameter $\lambda$.\\

Because of T-parity violation, the curves dip down at certain low
temperature that depend on the mass of the heavy photon and the
T-parity violation parameter $\lambda$. For an illustration, we
choose three typical values of lifetimes of heavy photon which
should be sufficiently long compared to the age of our universe,
otherwise it may contradict to the observation of the dark matter
density, as $10^{44}$, $10^{42}$ and $10^{40}$ $\textrm{GeV}^{-1}$
(about $2\times 10^{12}, 2\times 10^{10}$ and $2\times 10^{8}$
years) corresponding to the dimensionless T-parity violation
parameter $\lambda=3.00\times 10^{-23}$, $3.00\times 10^{-22}$,
$3.00\times 10^{-21}$ for $M_{A_{H}}=140$ GeV; $\lambda=2.72\times
10^{-23}$, $2.72\times 10^{-22}$, $2.72\times 10^{-21}$ for
$M_{A_{H}}=170$ GeV; $\lambda=2.08\times 10^{-23}$, $2.08\times
10^{-22}$, $2.08\times 10^{-21}$ for $M_{A_{H}}=290$ GeV
respectively, The three cases are shown in Fig. \ref{figure6}.
\\[\intextsep]
\begin{minipage}{\textwidth}
\centering \centering
\includegraphics[height=8cm]{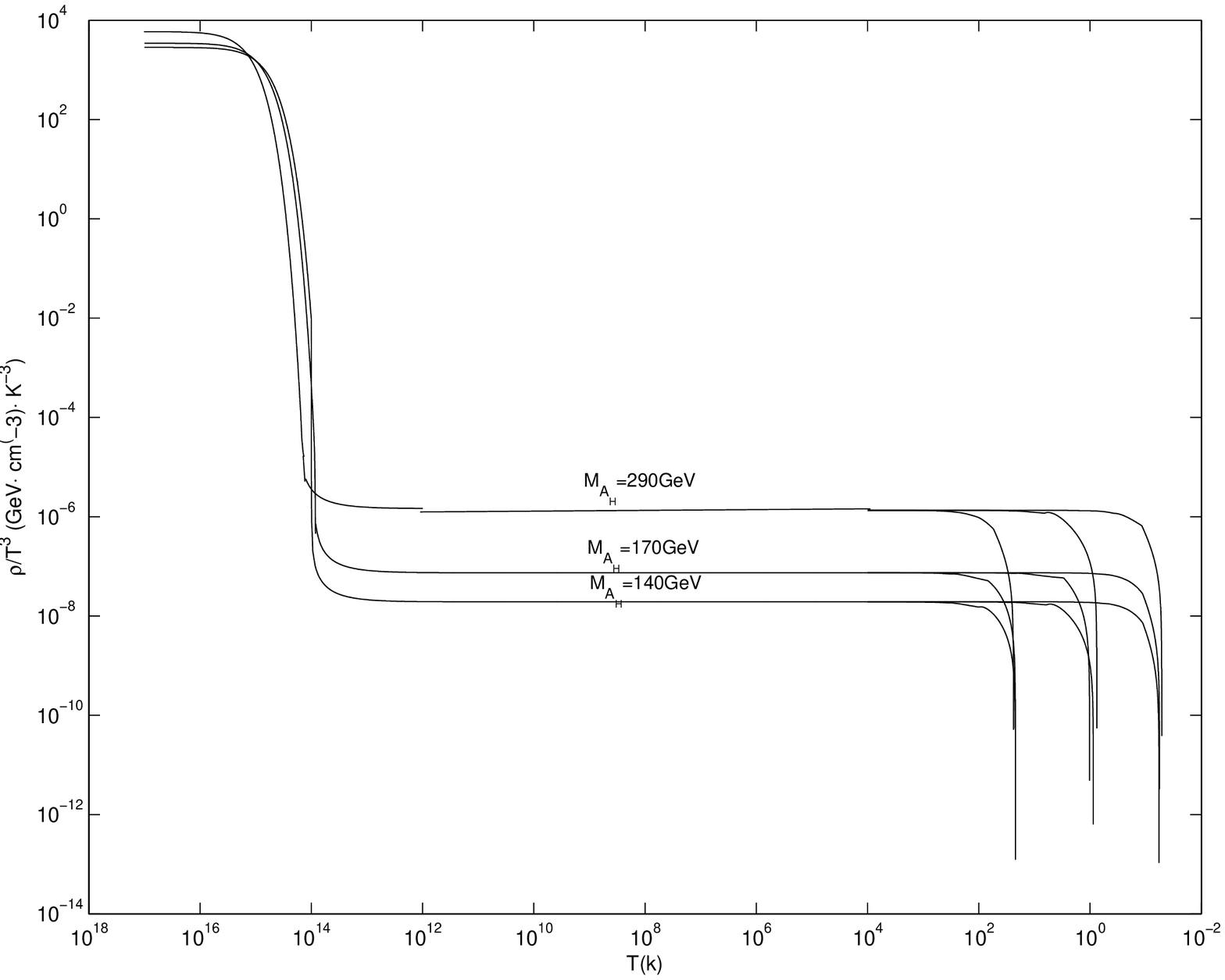}
\figcaption{ $\rho/T^{3}$ vs T. Different values for the heavy
photon mass $M_{A_{H}}$ and T-parity violation parameter $\lambda$
are taken.}\label{figure6}
\end{minipage}\\
\\[\intextsep]

By measuring the anisotropy, the density of CDM is determined as~\cite{PDG:2006}
\begin{eqnarray}\Omega _{dm} h^2 = 0.105_{ -
0.010}^{ + 0.007}~,\end{eqnarray} where $h$ is the Hubble constant
in units of 100 km/(s-Mpc) and we can get the range of present CDM
density $\rho_{CDM}$,
\begin{eqnarray}49.30~\textrm{eV}\cdot \textrm{cm}^{-3}\cdot
\textrm{K}^{-3}\leq \rho_{CDM}/T_{r}\leq 58.12~\textrm{eV}\cdot
\textrm{cm}^{-3}\cdot \textrm{K}^{-3}~.\end{eqnarray}\\

Obviously if the heavy photon is the unique constituent of CDM, it must
account for the whole CDM energy density, and the curves in Fig.
\ref{figure6} must pass through or intersect with the range  Eq.
(21).\\

The figure shows that there is only a narrow window for its mass to
be allowed by the measurements on the dark matter density. For
clarity, in Fig. \ref{figure7}, we show the relation between the
density of CDM, ($\rho/T^{3}$) in the flat region of curves in Fig.
\ref{figure4} and the mass of heavy photon $M_{A_{H}}$. The space
between the two horizontal lines corresponds to the observational
data with error tolerance. It is shown that without T-parity
violation, i.e. heavy photon is absolutely stable, only two regions
$133<M_{A_{H}}<135$ GeV and $167<M_{A_{H}}<169$ GeV can be
consistent with astrophysical data, if $135<M_{A_{H}}<167$ GeV,
there must be at least another kind of heavy particles contributing
to the rest CDM density. It is interesting to notice that as
$M_{A_H}$ changes, the cross section $<v\sigma>$ varies and near the
resonance shown in Fig. 3, it is proportional to
$(M_{A_H}^2/\Gamma_h^2)$ and becomes very large. Back to the master
equation Eq.(3), the solution of $\rho_0/T^3$ can reach almost zero
near the resonance. Therefore the two allowed ranges of $M_{A_H}$
are separated and the dependence of $\rho_0/T^3$ on $M_{A_H}$
is not monotonic.\\

\begin{minipage}{\textwidth}
\centering \centering
\includegraphics[height=8cm]{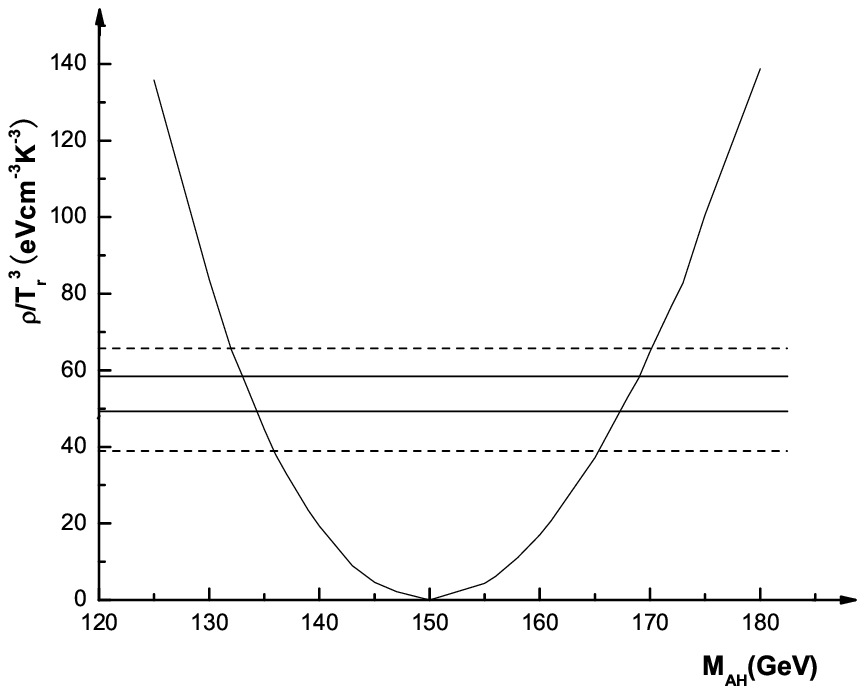}
\figcaption{Relation between the CDM density $\rho/T^{3}=n\times
M_{A_{H}}/T^{3}$ and the heavy photon mass $M_{A_{H}}$. The middle
region is allowed by the present astronomical observation.}
\label{figure7}
\end{minipage}\\
\\[\intextsep]

If T-parity violation indeed exists, the situation would be
different. For 1 $\sigma$ error tolerance, nearly all values of
$M_{A_{H}}$ except $135<M_{A_{H}}<167$ GeV is possible, because
$M_{A_{H}}$ can decay via T-parity violation into light SM
particles. However, on the other aspect, its relic density must pass
through the allowed region which is ranged by the solid lines (1
$\sigma$) or dashed lines (3 $\sigma$) of Fig. \ref{figure7} , that
is constrained by the present astronomical observation data. This
as well sets a restriction to the T-parity violation parameter $\lambda$
for different $M_{A_{H}}$.\\
\begin{minipage}{\textwidth}
\centering
\includegraphics[height=7.6cm]{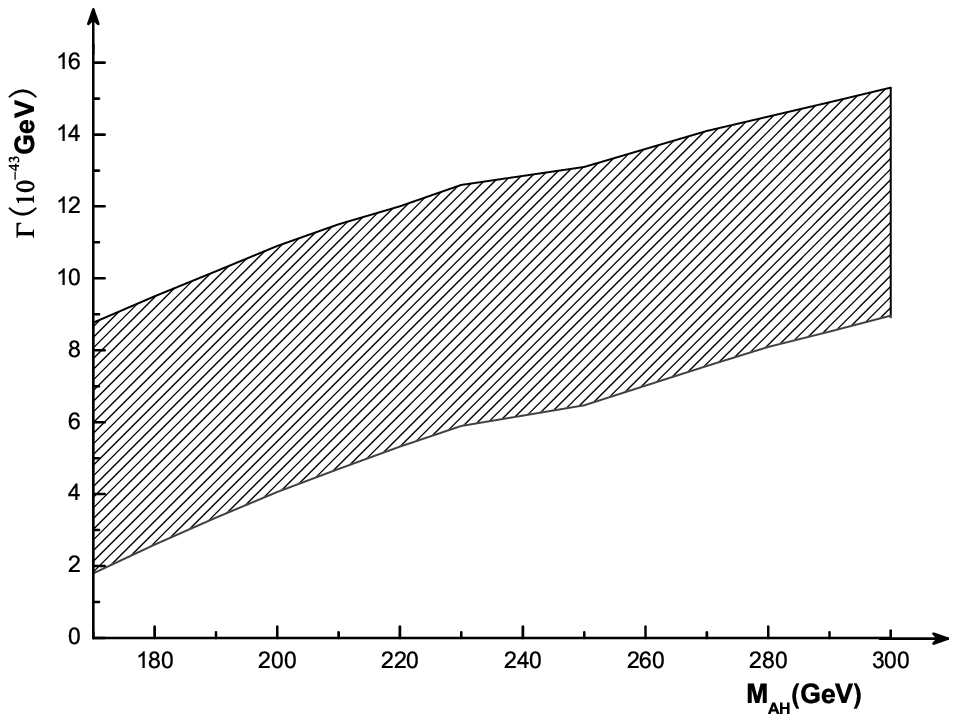}
\figcaption{The relation of $M_{A_{H}}$ and $\Gamma$ as
$M_{A_{H}}>$170
 GeV. The shaded area is allowed by the present
observation, with here we only consider the situation of
$M_{A_H}\geq 170$ GeV for an illustration in the $3\sigma$ error
tolerance.} \label{figure8}
\end{minipage}\\
\\[\intextsep]\\

To be explicit, we present the relation between $M_{A_{H}}$ and the
decay width $\lambda$ which is constrained by data, in Fig. 8.
Unfortunately, the relation cannot be written in a simple analytic
form, so that we illustrate it graphically. As
$M_{A_{H}}$ gets larger, the lifetime should be shorter, i.e. the
T-parity violation effect should be stronger (the violation parameter is
larger) accordingly. Thus if we assume $M_{A_{H}}$ to be a component
of CDM, the T-parity violation parameter is rigorously
constrained as it decays slowly to meet the present observation.\\

It is interesting to investigate how the initial conditions
influence the evolution, in other words, at very high temperature,
if we change the initial density of heavy photon, whether we can
obtain different relic density at present time. We choose three
different initial values at high temperatures and draw the evolution
curves in Fig. \ref{figure9}, the numerical results indicate that
the evolution is almost independent of the initial conditions. It
implies that the relic density is only related to the decoupling
temperature and mass of the CDM particles.\\
\begin{minipage}{\textwidth}
\centering \centering
\includegraphics[height=8cm]{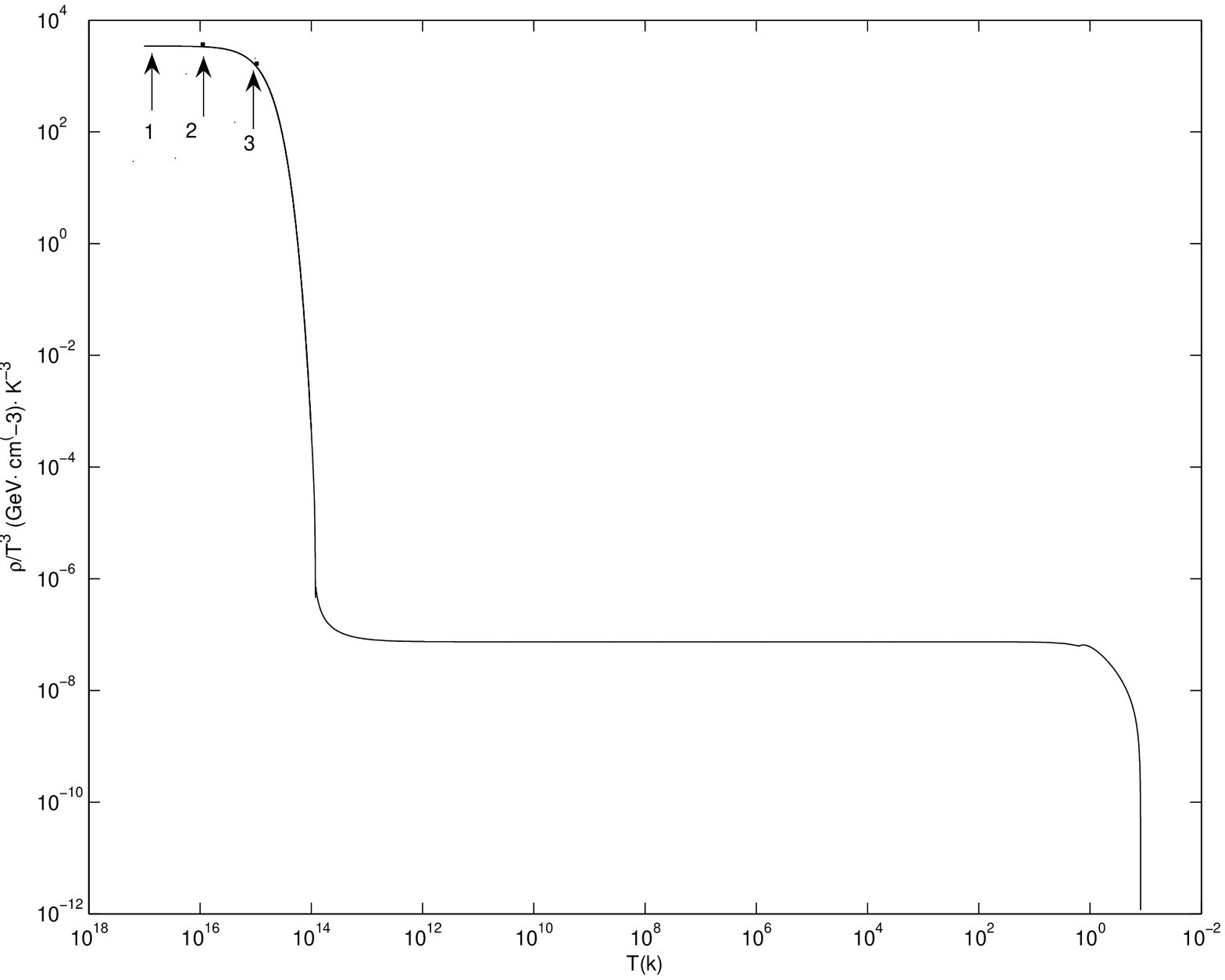}
\figcaption{$\rho/T^{3}$ on $T$ with various initial conditions.
Here $M_{A_{H}}$ is taken as 170 GeV.} \label{figure9}
\end{minipage}\\
\\[\intextsep]

\section*{III. Direct detection experiment } \hspace{0.3cm}\vspace{2mm}\\

There are many attempts to detect dark matter, either directly or
indirectly. In this work, we only concern the possibility of directly
detecting the dark matter flux.\\

In direct detection experiments one measures the dark matter flux
via elastic scatterings of the heavy photon off nuclei (proton). In
this work, we would theoretically evaluate the cross section in the
Littlest Higgs model and estimate probability of observing such
processes.\\

We suppose that the heavy photon is the only component of the CDM
flux, and then whether we can have sufficient data to find a clear
signal depends on the flux density and the dispersive velocity of
$A_{H}$ in the solar neighborhood and the coupling of heavy photon
with proton. This information about the concerned parameters is
obtained by studying the relic
density of CDM and that procedure was carried out in previous
sections, we then evaluate the rate of events expected in an earth
experiment (i.e. LTP-proton scattering events) per unit time, per
unit detector material mass. The events number is approximately
given by
\begin{eqnarray}
\Delta n = N<\sigma> n\overline{v}\Delta t\varepsilon ,
\end{eqnarray}
where $N$ is the number of target protons in the detector, $n=\rho_0
/M_{A_{H}}$ ( $\rho_0=0.3~GeV/cm^{3}$ ) ~\cite{PDG:2006} is the
local heavy photon number density and $<\sigma v> \thickapprox
<\sigma> \overline{v}$ is the cross section for the scattering which
is averaged over the relative velocity of $A_{H}$ in the lab frame,
$\overline{v}=<v^{2}>^{\frac{1}{2}}$ is the dispersive velocity of
dark-matter particles, $\Delta t$ is the time of the detection,
$\varepsilon$ is the detection efficiency. In order to get a
plausible numerical result, we select 1000 kg water, thus
$N=N_{A}10^{6}g/(18g/mol)\times10$, the local energy density of dark
matter $A_{H} (\rho_{0})$ is $0.3~GeV/cm^{3}$ \cite{PDG:2006}, the
mass of $A_{H} (M_{A_{H}})$, is $175~GeV$ and $\Delta t$ is set to
be
one year.\\

The formula of the average cross section $<\sigma>$ is shown as below:
\begin{eqnarray}
 < \sigma (v_1 ) > = \int\limits_0^\infty {f(v_1 )} \sigma (v_1
 )d^3v_1 ,
\end{eqnarray}
where $v_1=|\overrightarrow{v_1}|$ is the velocity of the $A_H$ in
the lab frame where the target proton is at rest, and
$$ f(v_1 )d^3v_1 = 4\pi e^{ - v_1^2 / v_0^2 } /
(\pi ^{3 / 2}v_0^3 )v_1^2 dv_1 $$~\cite{Gerard:1996} is the local
Maxwellian velocity distribution, by
$\mathrm{\overline{v}=<v^{2}>^{\frac{1}{2}}=270~km/s}$~\cite{Gerard:1996,Bertone:2004pz}
, we know $v_0^{2}=5.40789\times10^{-7}c^{2}$.\\

The expression for the cross section is a bit more complicated
\begin{eqnarray}\sigma (v_1 ) = g(v_1 )[\arctan ( - \frac{M_H
}{\Gamma _H }) - \arctan ( - \frac{4p_1^2 - M_{_H }^2 }{\Gamma _H
})]~,\end{eqnarray} where we take the Higgs mass
$\mathrm{M_H=300~GeV}$, $\Gamma_H$ is the decay width of Higgs,
$p_1=|\overrightarrow{p_1}|=M_p v_1 / (M_{A_H } + M_p )$ is the
momentum of $A_H$ in the CM frame and
\begin{eqnarray}
g(v_1 ) = \frac{{g}'^4M_p^4 }{192\pi p_1^2 M_H \Gamma _H (\sqrt
{p_1^2 + M_{A_H }^2 } + \sqrt {p_1^2 + M_p^2 } )^2}~.
\end{eqnarray}\\

The Feynman diagram of heavy photon scattering off proton is shown
in Fig. \ref{feyn3}.\\
\\[\intextsep]
\begin{minipage}{\textwidth}
\centering
\includegraphics[height=3cm,angle=0]{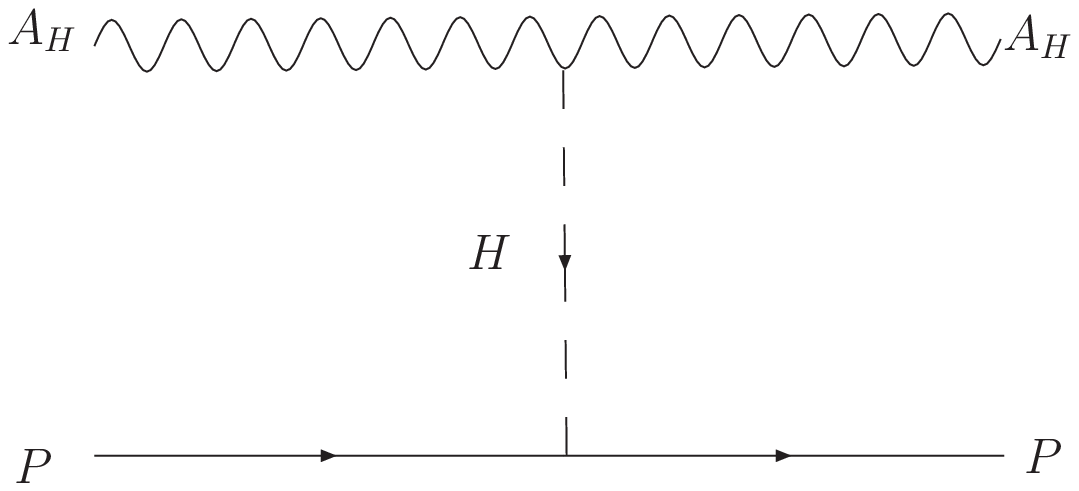} \figcaption{Feynman
 diagrams for the directly detecting the dark matter ($A_{H}$).} \label{feyn3}
\end{minipage}\\
\\[\intextsep]

Through equations (23)-(26), we know that if the detection
efficiency $\varepsilon$ is about 0.1, during one year period, in
1000 kg water, there can only be $5.8\times 10^{-16}$ signals. If we
want to obtain 58 signals, we need at least $10^{17}$ tons water,
that is to say, we need to build a 10 m high, $10^{8}~m$ wide and
$10^{8}~m$ long pool, which sounds unreasonable. At the same time,
Unfortunately, due to the extremely small interaction rates and the
small energy transferred to the recoiled nucleus that if the dark
matter particle is heavy photon, it is almost impossible
to distinguish the events from backgrounds.\\

\section*{IV. Conclusion } \hspace{0.3cm}\vspace{2mm}\\

The purpose of measuring dark matter flux is not only to measure the
flux itself, but also to explore new physics beyond the standard
model. Indeed, as indicated in all literature, the CDM particles
cannot be SM ones, therefore study on the CDM flux may provide
valuable information about the new physics. It is lucky that the LHC
will begin running in 2008, one may expect to find signals about new
physics beyond the SM from a great amount of  available data. Thus
the cosmology constraint on the new physics will stand as
complimentary to the accelerator experiments.\\

There have been many proposals about the CDM candidates in
literature, such as the SUSY particles neutralino~\cite{JHEP:2005},
sneutrino\cite{Arina:2007tm,Asaka:2007zz,Asaka:2006fs},
axino\cite{Hannestad:2007dd,Carosi:2007zz,Kinion:2004ab},
technicolor-particles\cite{Hur:2007uz,Gudnason:2006yj,Gudnason:2006ug},
heavy photon in the Littlest Higgs
model\cite{Perelstein:2006bq,Birkedal:2006fz,Martin:2006ss} and even
unparticles\cite{Deshpande:2007jy,Kikuchi:2007az}, it is hoped to
find an efficient way to distinguish signals originating from
different models. In fact, by solving the Boltzman-Lee-Weinberg
equation and making the results to be consistent with the
observational data, one can set constraints on the model parameters,
such as the mass of the CDM candidate particle and its coupling to
the SM particles if its decay is allowed. This estimation will
provide information for the accelerator experiments which are
searching for such particles on earth. Moreover, the probability of
directly detecting of the dark matter particles on earth is another
way to distinguish the various models. Namely, even though all the
the models can be consistent with the astronomical observation on
the relic dark matter, they may have completely different
probabilities to be caught by the detector on earth. By measuring
the dark matter flux, we can rule out some possible candidates and
then combining the data of LHC, one may eventually determine the
identity of the new physics particle for
CDM.\\

In this work, a possible Dark Matter candidate -- heavy photon in
the Littlest Higgs model with T-parity has been discussed. We
observe that different Higgs and heavy photon masses can result in
totally distinctive average value of the heavy photon
pair-annihilation cross section, which determines the time evolution
of our universe. By employing a sample Higgs mass $M_{H}=300$ GeV
without losing generality, we present the relationship between the
density of cold dark matter $\rho/T^3$ and the temperature $T$ for
different stages. Thanks to the newly available data and observation
of astrophysics, we can set a stronger constraint on the mass of
heavy photon. Only two ranges $133<M_{A_{H}}<135$ GeV and
$167<M_{A_{H}}<169$ GeV are allowed without T-parity violation ,
whereas there must be at least another kind of heavy particle
contributing to the rest CDM energy density for the range
$135<M_{A_{H}}<167$ GeV.  Outside the three ranges, the Littlest
model cannot meet the data. If dark matter particles are found to
fall in the regions beyond 132 to 170 GeV (of course with some
errors), then the Littlest Higgs model without
 T-parity violation would be ruled out by the cosmology data. If
 T-parity violation was considered, the data would set rigorous constraints
  on T-parity violation  parameter $\lambda$. We also calculate the
possibility of directly detecting of this type of Dark Matter, i.e.
heavy photon on the earth. Our conclusion is that it is impossible
to observe any signal due to small interaction rates and severe
backgrounds. In other words, if heavy photon is the CDM particle, it
can easily evade our detection on earth. Since different CDM
particles have different probabilities to be detected by earth
detectors, our results offer a way to distinguish the Littlest Higgs
model from others. If our detectors on earth do not catch any dark
matter flux particles, the Littlest Higgs model may survive,
otherwise, one can erase such a model from our model
collections.\\
\\
\noindent {\bf Acknowledgements}:\\

This work is partially supported by the National Natural Science
Foundation of China (NNSF).\\

\end{document}